\documentclass[aps, prb, twocolumn, superscriptaddress]{revtex4}
 \usepackage{amsmath, amssymb}
 \usepackage{bm, color}

\usepackage{graphicx}
\usepackage{hyperref}

\usepackage{natbib}

\newcommand{\be}{\begin{equation}}
\newcommand{\ee}{\end{equation}}

\begin{document}

\title{Comment on ``Quantum oscillations in nanofabricated rings of spin-triplet superconductor Sr$_{\bf 2}$RuO$_{\bf 4}$''}

\author{V.~Vakaryuk}
\email[]{vakaryuk@gmail.com}
\affiliation{Materials Science Division, Argonne National Laboratory, Argonne, Illinois 60439, USA}

\author{K.~Roberts}
\affiliation{Department of Physics, University of Illinois at Urbana-Champaign, Urbana, Illinois 61801, USA }

\author{D.G.~Ferguson}
\affiliation{Department of Physics and Astronomy, Northwestern University, Evanston, Illinois 60208, USA}

\author{J.~Jang}
\affiliation{Department of Physics, University of Illinois at Urbana-Champaign, Urbana, Illinois 61801, USA }

\author{R.~Budakian}
\affiliation{Department of Physics, University of Illinois at Urbana-Champaign, Urbana, Illinois 61801, USA }

\author{S.B.~Chung}
\affiliation{Department of Physics, McCullough Building, Stanford University, Stanford, California 94305, USA}



\maketitle

Recently Jang \textit{et al.}\cite{Jang:2011} reported the observation of half-height  magnetization steps (``half-steps'') in cantilever magnetometry measurements of mesoscopic annular $\rm Sr_2 RuO_4$ particles. Such magnetization features were interpreted as the presence of half-quantum vortices (HQVs)\cite{Footnote1}. In an attempt to examine our findings, very recently Cai \textit{at el.}\cite{YL:2012} have performed magnetotransport measurements  of micron-size rings fabricated from small $\rm Sr_2 RuO_4$  crystals.  While fabrication of such samples and subsequent verification of our findings is highly desirable, we would like to point out that,  at the current state of affairs, the direct comparison is incomplete partly due to the fact that the measurements of Ref.~\onlinecite{YL:2012} were lacking an important ingredient -- the in-plane magnetic field. We would also like to offer clarification on few questionable statements made in Ref.~\onlinecite{YL:2012}.

Cai \textit{et al.} raised a concern that the field periodicity $\Delta H$ of the magnetic response  in our measurements disagrees with the apparent geometry of our samples\cite{Footnote2}. This concern is unfounded. The criticism of Cai \textit{et al.} is presumably based on the use of the inner hole radius to estimate $\Delta H$. However if the penetration depth $\lambda$ is comparable with all relevant dimensions of the sample (as is the case in our measurements) a more elaborate treatment is needed. To show this we use a model of an infinite superconducting cylinder with the inner and outer radii $R_1$ and $R_2$ respectively oriented along the field\cite{Arutunian:1983}.  The periodicity of the magnetic response is twice as large as the first fluxoid entry field which, for a cylinder of arbitrary dimensions, is given by Eq.~(27) of Ref.~\onlinecite{Vakaryuk:2011PRB}.
Taking $R_1=390\,$nm, $R_2=850\,$nm and $\lambda = 200\,$nm (as appropriate for our sample and the temperature of our measurements) we obtain $\Delta H \approx 21\,$G. Given the irregular shape of our samples and uncertainty in the knowledge of $\lambda$, this result is in a very reasonable agreement with the experimentally observed value of 16\,G.

Since the comparison with an infinite geometry might be  arguable, we have also performed finite element numerical analysis on realistic sample geometries with the help of commercially available software package \textsc{comsol}. We have simulated the field response of our samples using Ginzburg-Landau theory with material parameters appropriate for $\rm Sr_2RuO_4$ (including out-of-plane anisotropy)\cite{Kevin:2011}. We have found that truncating the cylinder to a slab with a height 300\,nm (other parameters as quoted above) yields $\Delta H=  19.8  \,$G  thus confirming our earlier conclusions.


%
\begin{figure}[b]
\centering
	\includegraphics[scale=.45]{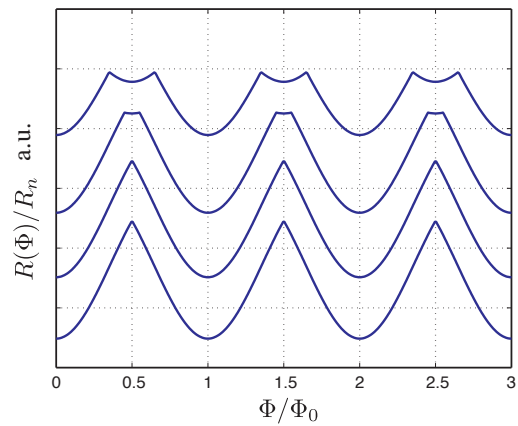}
	\caption{\label{MR} Magnetoresistance of a superconducting ring close to the transition temperature as a function of the applied flux for several values of the in-plane field: $H_x=0$, $H_x = H_{x0}$, $H_x = 2H_{x0}$ and $H_x= 4 H_{x0}$ from the bottom to the top. Here $H_{x0}$ is defined as a field below which the HQV is thermodynamically unstable.  The signatures of the HQVs are seen only for $H_x > H_{x0}$ and appear as a splitting of the magnetoresistance maxima. The curves are offset for clarity.} 
\end{figure}

We now point out a crucial difference between the experimental setups of Refs.~\onlinecite{Jang:2011} and \onlinecite{YL:2012} which make the direct comparison of their results incomplete. In Ref.~\onlinecite{Jang:2011} it was found that in the thermodynamic equilibrium half-steps appear \emph{only} in the presence of an in-plane magnetic field (magnetic field which lies in the ab-plane of the crystal, denoted hereafter as $H_x$). This field is additional to the c-axis field which is responsible for the measured component of magnetization\cite{Footnote3}. The half-steps appear when the magnitude of $H_x$ exceeds some  sample and temperature dependent value $H_{x0}$. Typically values of $H_{x0}$ lie in the range 10--100\,G. While the origin of the in-plane field coupling remains unclear, there do exist a few theoretical models which could account for it \cite{Vakaryuk:2009, Sigrist:2011}.

In the magnetotransport measurements of Ref.~\onlinecite{YL:2012} no in-plane field was applied. Given our earlier findings, under such conditions signatures of HQVs are not expected to appear. Cai \textit{et al.} motivate their measurement by referring to our field sweep data taken at $H_x=0$. At low temperatures magnetization measured under such conditions show significant hysteresis and, occasionally, half-height steps.  It must be emphasized however that such features correspond to \emph{metastable} states and are unlikely to be observed in reversible transport measurements.

To demonstrate this point we calculate the magnetoresistance of a superconducting ring close to the transition temperature by adapting a method developed by Sochnikov \textit{et al.}\cite{Sochnikov:2010} to a material which is capable of supporting HQVs\cite{Vakaryuk:2011PRL}.  Results of the calculation are shown on Fig.~\ref{MR} for several values of $H_x$.  The signatures of HQV appear only when $H_x > H_{x0}$ and correspond to the splitting of the magnetoresistance maxima. Contrary to the  statement made in Ref.~\onlinecite{YL:2012}, in our analysis the appearance of HQVs in magnetotransport does not break the $\Phi_0$ periodicity. This statement also holds for equilibrium magnetization measurements and is well understood in terms of the stability diagram for HQVs and regular vortices.

We also note in passing that, because of configurational entropy, HQVs in the bulk\cite{Chung:2010} or in an assemble of connected rings (network arrays)\cite{Sochnikov:2012} might be stabilized and hence seen in transport measurements even when the stability of a single HQV is not expected.

VV would like to thank Ying Liu for useful discussions.


\begin{thebibliography}{11}
\bibitem{Footnote1} In an annular geometry technically more appropriate term would be ``half-integer fluxoids'' which describes the half-integer winding of the phase coupled to the charge current  around the hole. We will colloquially refer to half-integer states as HQVs.
\bibitem{Footnote2} In general the field dependence of the magnetic response has a non-periodic Meissner component (linear at small applied fields) and a periodic vortex component (with period $\Delta H$) which is due to the fluxoid quantization.
\end{thebibliography}

\end{document}